%% file: main.tex
\documentclass[copyright,creativecommons]{eptcs}
\providecommand{\event}{DCDP 2010} 
\usepackage{breakurl}             
\usepackage{acronym}
\usepackage{verbatim}
\usepackage{fixltx2e}
\usepackage{subfigure}
\usepackage{graphicx}

\title{On Secure Workflow Decentralisation on the Internet}

\author{Petteri Kaskenpalo
\institute{AUT University\\ Auckland, New Zealand}
\institute{School of Computing and Mathematical Sciences}
\email{Petteri.Kaskenpalo@aut.ac.nz}
}
\def\titlerunning{On Secure Workflow Decentralisation on the Internet}
\def\authorrunning{P. Kaskenpalo}
\begin{document}
\maketitle

\begin{abstract}
Decentralised workflow management systems are a new research area, where most work to-date has focused on the system's overall architecture. As little attention has been given to the security aspects in such systems, we follow a security driven approach, and consider, from the perspective of available \emph{security building blocks}, how security can be implemented and what new opportunities are presented when empowering the decentralised environment with modern distributed security protocols.

Our research is motivated by a more general question of how to combine the positive enablers that email exchange enjoys, with the general benefits of workflow systems, and more specifically with the benefits that can be introduced in a decentralised environment. This aims to equip email users with a set of tools to manage the semantics of a message exchange, contents, participants and their roles in the exchange in an environment that provides inherent assurances of security and privacy.

This work is based on a survey of contemporary distributed security protocols, and considers how these protocols could be used in implementing a distributed workflow management system with decentralised control . We review a set of these protocols, focusing on the required message sequences in reviewing the protocols, and discuss how these security protocols provide the foundations for implementing core control-flow, data, and resource patterns in a distributed workflow environment.
\end{abstract}


%

\acrodef{YAWL}[YAWL]{Yet Another Workflow Language}

\section{Introduction}
The research area of distributed and decentralised workflow management is still an immature field with many open problems. A few distributed and even fewer decentralised workflow management systems have been discussed in the literature. L. Guo et al. \cite{Guo2008} provide an overview of the existing work, including those of \cite{Fakas2004,Yan2004,Guo2006}. As these earlier works do not discuss how security protocols could be used to address decentralised workflow related problems, we aim to provide a starting point with this article.

We were prompted to look into this area by the general lack of progress in responding to the needs of email users. While email is used to run various workflow process like activities, users are provided with little support in managing these. We believe that decentralised workflow management research will help in addressing this problem, and will return to this topic along our more generic discussion.

This work is based on a survey of contemporary distributed security protocols, and considers how these protocols could be used in implementing a distributed workflow management system with decentralised control. This article makes the following contributions. Firstly we review a set of contemporary distributed security protocols, focusing on the required message sequences in reviewing the protocols, and leaving out the cryptographic algorithms required to construct and process the message contents. This makes the protocols more accessible without expert knowledge in cryptography and mathematics, and enables us to focus on required interactions and involved participants. Secondly we discuss how these security protocols can provide the foundations for implementing secure core workflow control, data and resource patterns in a distributed environment. 

The advancement of secure distributed, decentralised processing requires well-understood building blocks that can be used in modelling processes at a higher abstraction level. We approach this goal from the angle of contemporary security protocols aimed at solving generic problems in distributed collaboration, and consider how these solutions can be utilised in the area of decentralised workflow management. In addition to addressing known problems in a new environment, we also suggest that these enable new workflow functionality relating to privacy, anonymity and group based identity.



\subsection{Motivation}

Email is the most used communication tool in business today and its use has been extended far beyond its original purpose of simple information exchange \cite{Lusk2006}. One reason for this success is the asynchronous and distributed nature of the email message exchange, which by definition retains control over the communication, its contents, and time of interaction with the end-users.

The success of email has certainly also been its Achilles heel, as the tool is stretched to areas it was not designed for. Examples of everyday email uses now include: information management; task and time management; multi-party activity co-ordination; distributed decision making; negotiations; and voting as a discussion facilitator, to name a few. In many ways, email is being used to manage complex, collaborative process activities. This development has, of course, been further accelerated with the ability to access emails everywhere via mobile devices.

However, considering the relatively long development record of messaging tools, it is striking how its very success has led to its many problems; difficulties in dealing with message overflow, unsolicited messages, and message linking, archiving and recovery. Inefficiencies can be considerable, as the end-users do not have enough time to manage these intermingled, unstructured message flows.

Positive enablers, like flexibility and end-user-centredness of email systems, remain in distinct contrast with today's workflow management systems, which require centrally configured process rules and centrally managed process execution. They give little freedom to the users with regards self-organisation, task scheduling, and little inherent assurance of security and privacy due to their centrally managed nature. We suggest that it is possible to combine the end-user-centredness of email with the benefits of workflow systems in a decentralised email platform that would equip its users with a set of tools to manage the semantics of a message exchange, contents, participants and their roles in the exchange.

We have suggested the \emph{Service Oriented Email} (SOE)~\cite{Kaskenpalo2009} concept to address these challenges in a structured and controlled manner, and at the same time provide an organisational approach for solving email-related workflow process problems. Our research advances the notion of semantic email as introduced by McDowell et al. \cite{McDowell2004}, and proposes a scalable distributed semantic email platform for automating complex multiparty update, query, resolution and process activities without exposing individuals beyond customary expectations for email privacy and control.

The key construct in the SOE concept is the ability of end-users to publish email message interfaces that define the service they provide and what they require for execution. By chaining these message interfaces, end users could be enabled to build complex distributed activities. The users should also be able to manage the organisational or social structures in which they participate by creating groups and managing group participants, and by linking a group with higher level hierarchical units. This bears a similarity with the paradigm of multiagent systems (\emph{MAS}), where the provided tools for cooperation and coordination between agents are probably its most important contribution. However, rather than applying these principles to autonomous agents, we suggest to equip end-users with similar tools to assist them.

To achieve this goal, we require the support of distributed security protocols that provide the core security services for the messaging platform, and suitable modelling constructs for defining and controlling the message exchanges. However, we do not limit our discussion in this article only to the SOE environment, but take a general view of the decentralised workflow enactment problem.

We accept that all of the discussed protocols assume the existence of a basic public key infrastructure and a trusted source to assign keys to participants. However, this task could be handled by multiple participants, depending on their existing connections in the hierarchy, for example, in a company setting where all users could be set up by the  human resources department, and the subsequent management of group memberships could be managed amongst the users. An alternative approach would be to provide a group key scheme establishment service that would facilitate the initial group key setup, similar to current Internet certificate providers, but that would not participate in the subsequent key management.

\subsection{State of workflow systems}
Workflow and process-driven systems are, of course, a well-studied area of computing. The work of Aalst et al. in \emph{Workflow Patterns} \cite{VanDerAalst2003} provides a comprehensive review of workflow building blocks, and discusses the workflow specifications from a number of different perspectives; namely the controlflow, data, resource and operational perspectives. They make a distinction between twenty different control-flow patterns. The work of Aalst et al. in \cite{VanDerAalst2003} has been further advanced by Russell in \cite{Russell2007} by analysing patterns supported by various workflow management systems, and by theoretical analysis of possible patterns not yet considered. His work suggests an additional 23 controlflow patterns and many additional patterns in the other categories.

While the works of Aalst et al. and Russell aim to \emph{'identify comprehensive workflow functionality'} and to \emph{'provide the basis for an in-depth comparison of a number of commercially available workflow management systems'} \cite{Aalst2003}, they also provide an invaluable foundation for further study of workflow management systems, especially when moving from centrally controlled, non-distributed workflow management systems towards workflow systems that are distributed and controlled in a truly decentralised manner. We build on this earlier work and consider how these patterns can be implemented in a decentralised environment with the help of the security protocols, and what additional patterns could be suggested for this type of system and operating environment. We discuss examples of these in section \ref{distributedpatterns}.

\subsection{Roadmap}

The rest of this article has the following structure. In section \ref{securityprotocols} we describe contemporary distributed security protocols that are essential in extending the application of workflow systems to a distributed processing model, and outline the message exchanges required by these protocols using sequence diagrams.
In section \ref{distributedpatterns} we discuss the implementation of distributed, decentralised workflows by analysing a selected set of recognised workflow patterns. 

\input{securityprotocols}

\input{otherprotocols}

\input{distributedpatterns}

\input{distributedwfsim}

\section{Conclusion}
Distributed security protocols for group communications provide a mechanism for implementing data confidentiality, authenticity, non-repudiation and privacy for collaborative workflow environments. They enable a flexible approach where these security objectives can be met at different levels of granularity. We have reviewed a number of contemporary distributed security protocols, and discussed how these can be utilised to implement the security functionality expected of centralised workflow systems, as well as new functionality enabled by the very nature of de-centralisation. For example, from the perspective of privacy the distributed environment offers further benefits in the form of participant anonymity within a group as well as being able to act on behalf of a group rather than an individual participant.

\bibliographystyle{eptcs}
\bibliography{references}
\end{document}

%% file: securityprotocols.tex
\section{Security Protocols}
\label{securityprotocols}

Over the years many promising distributed security protocols have been proposed for handling a multitude of situations requiring participant collaboration in achieving a shared goal. We introduce the following protocols for building distributed workflow systems:
\begin{description}
  \item[A group signature scheme]Enables a group to sign messages in a way that a subset or all members of the group are required to provide their signature for the group signature to be valid. Later no member can deny their participation in providing this group signature.
  \item[A group safe scheme] Distributes a secret between the members in a way that no member can reproduce the data without the others. If one member loses their share of the encryption key or their personal master key, this does not compromise the shared secret.
  \item[A group coin toss protocol] Enables the participants to make a choice, and commit to these without being able to change their selection between the commitment and revealing the choice. This has a number of uses, and can be used for example in implementing an anonymous group communication protocol.
  \item[A distributed Byzantine agreement protocol] Provides a mechanism for agreeing on proposed values between the participants. This can be used to implement distributed data replication schemes and distributed notice board services.
  \item[A Conference key protocol] Enables the confidentiality of message exchange in a group. Lastly, anonymous broadcast protocols enable a group member to broadcast a message without revealing the sender's (or receivers') identity.
\end{description}

\begin{figure}[htbp]
\begin{center}
\includegraphics[scale=.5]{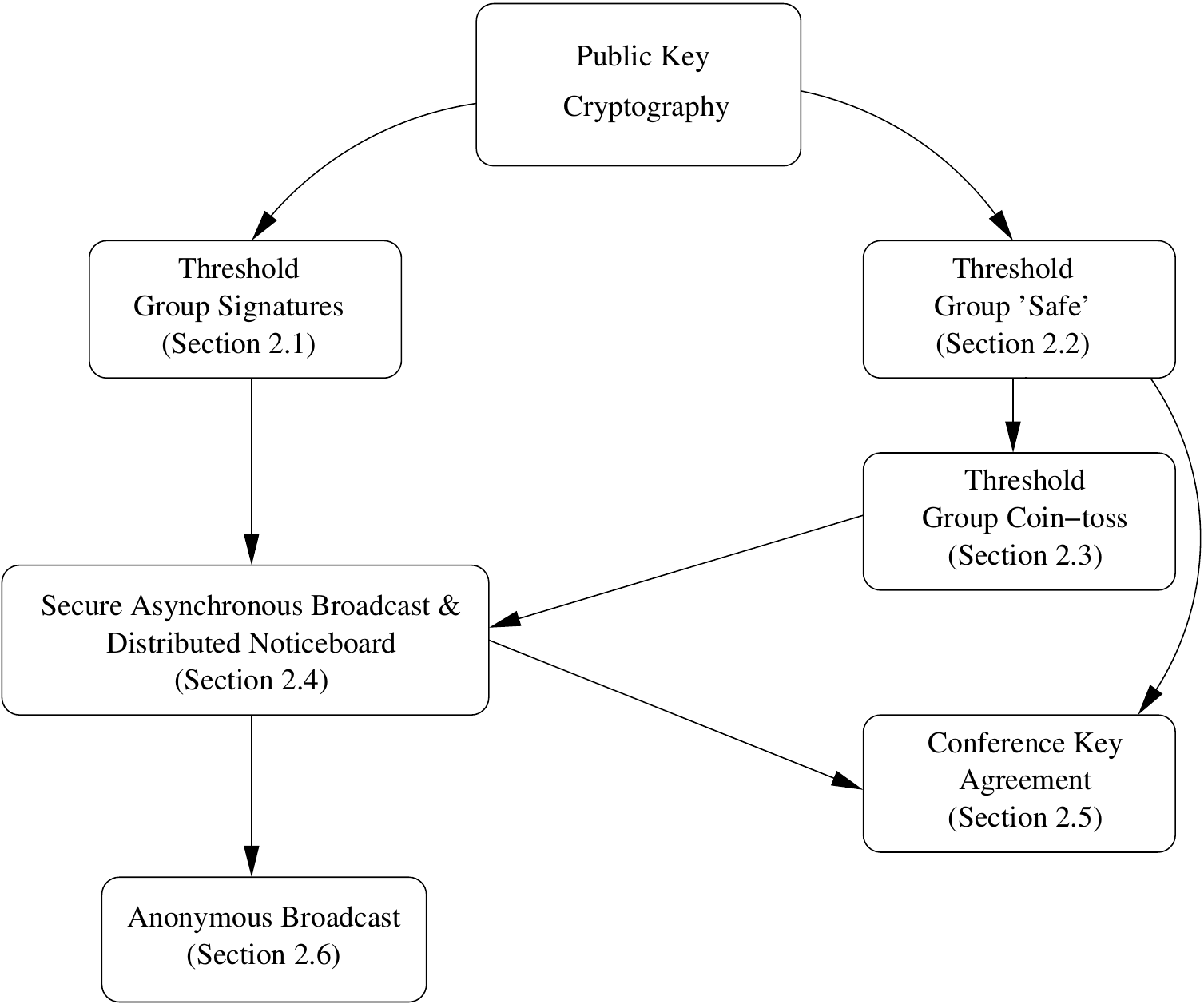}
\caption{Overview of reviewed security protocols and their interdependencies}
\label{fig:secprotocols}
\end{center}
\end{figure}

The relationships between these protocols are depicted in figure \ref{fig:secprotocols}. For each of the protocols we will describe the problem they have been created to solve, provide an overview of the solution, show the message exchange between the participants, and give an example of a situation where the protocol can be used. We focus only on showing the message exchange of the protocols rather than discussing the cryptographic algorithms required to construct and process the message contents, as our focus is on discussing how these protocols can be implemented in a workflow environment (see section \ref{distributedpatterns}).

We are only able to discuss a small set of useful security protocols here. New developments in the security protocol field will continue to provide us with tools that will enable us to perform activities in a more efficient way, or enable us to do entirely new things. For instance, we have not discussed \emph{Homomorphic encryption} protocols \cite{Fontaine2007}, where an activity can be performed on encrypted content and later reveal the result by decrypting the content, for example, we can add numbers to an encrypted value, without having knowledge of the value. However, if the encrypted sum is decrypted, the actual sum of the two is revealed. Homomorphic encryption has been used in creating secure voting systems, collision-resistant hash functions, and private information retrieval schemes.

\input{thresholdsignatures}

\input{thresholddatascheme}

\input{thresholdcointoss}

\input{distsecagreement}

\input{confkey}

\input{anonymousbroadcast}


%% file: thresholdsignatures.tex
\subsection{Threshold group signature}
\label{thresholdsignatures}

In a group signing situation each of the participating signatories provides an individual signature; however, the overall signature is not valid unless the combined signature of all signatories is valid. A group signature scheme enables a group of participants to demonstrate the authenticity of messages by enabling a verifier to construct a composite signature from the \emph{signature shares} provided by each signing party, and by enabling this composite signature to be verified by one public key pre-assigned to the group of signatories. In a \emph{(k,l)-threshold signature} scheme at least \emph{k} of the \emph{l} parties are required for signing.

\begin{figure}[htbp]
\begin{center}
\includegraphics[scale=.35]{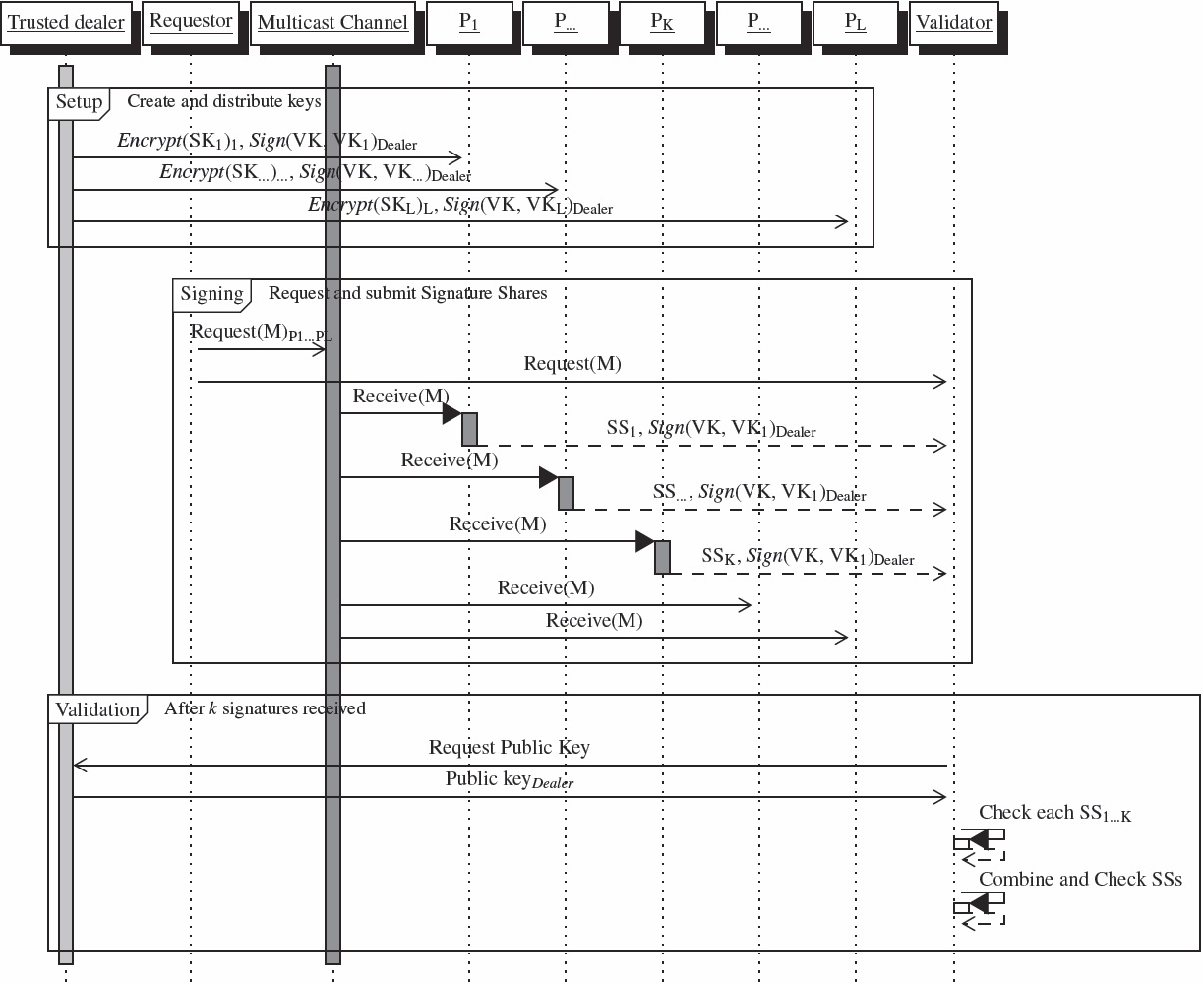}
\caption{Message exchange in the threshold signature protocol, where \emph{SK = Key Share, VK = Verification Key, M = Message to Sign, and SS = Signature Share}}
\label{thresholdsignaturesequences}
\end{center}
\end{figure}

Figure \ref{thresholdsignaturesequences} depicts the participating parties and the key message exchange for a threshold signature protocol described in \cite{Shoup2000}. In the first stage, the trusted dealer provides each participant with their key shares. In the signing stage, each participant that agrees to sign the document forwards their individual signature shares for validation of the group signature. A group signature scheme could be used to sign an agreement by multiple parties over an asynchronous messaging platform. Threshold signatures are required in section \ref{distnoticeboard} to implement a security asynchronous broadcast and a distributed noticeboard protocol. These applications also require a distributed threshold coin-tossing protocol, which we will review in Section \ref{thresholdcointoss}.

%% file: thresholddatascheme.tex
\subsection{Threshold group \emph{'safe'}}
\label{thresholddata}
How can some secret data be shared among a number of participants in a way that no one will have control of the data over the other group members? We could store multiple copies of the secret, but this could result in a leak if one of the members is compromised. Similarly to the threshold signatures, some secret data \emph{D} can be divided into \emph{n} pieces in such a way that for any number \emph{k} or more \emph{D\textsubscript{i}} pieces  will make \emph{D} easily computable, but any less than k pieces will leave \emph{D} incomputable \cite{Shamir1979}. In other words, \emph{k} sets the threshold of how many parties are required to co-operate to recreate the data. This threshold can be set when the protocol is initiated for each data item.

This protocol enables the implementation of a distributed \emph{safe}, where no single party owns the entire set of data, and where \emph{k} members are required to compose the data together. In other words, a group of mutually suspicious participants must cooperate. This is useful, for example, in storing and sharing secret keys among group members. 

\begin{figure}[htbp]
\begin{center}
\includegraphics[scale=.35]{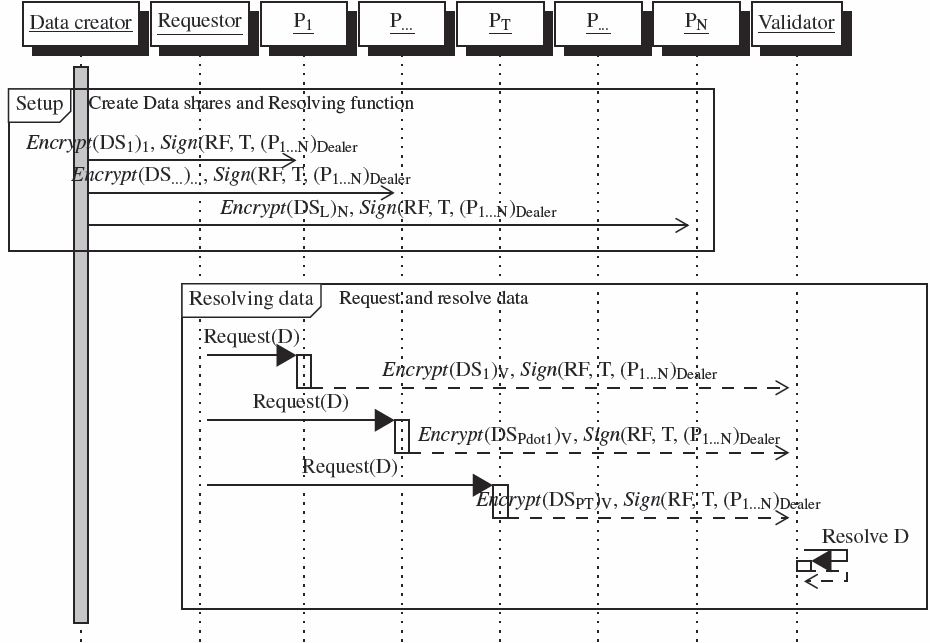}
\caption{Message exchange in the threshold group safe protocol, where \emph{DS = Data Share, RF = Resolving Function, and T = Threshold}}
\label{fig:thresholddata}
\end{center}
\end{figure}

Figure \ref{fig:thresholddata} show the protocol messages and participants. An interesting example of the use of the protocol is the implementation of hierarchical organisation structures by, for example, giving the company's president three values of \emph{D\textsubscript{i}}, each vice-president two values of \emph{D\textsubscript{i}}, and each executive one value of \emph{D\textsubscript{i}}. In a threshold data safety scheme that requires threeout of \emph{N} data shares, this arrangement enables documents to be signed by the president alone, by two of the vice-presidents or by three of the executives. This secret-sharing protocol provides a convenient building block for a distributed coin-tossing protocol, which we will discuss in the next section.

%% file: thresholdcointoss.tex
\subsection{Threshold group coin toss}
\label{thresholdcointoss}
The problem is how to toss a coin between two parties who are not physically together, and prevent either party from lying about the result. The problem can be solved if the coin toss result can be stored in a way that can not be altered by the party that tossed the coin. The threshold group safe scheme discussed in the previous section is a suitable tool for this purpose, and Cachin et al. \cite{Cachin2005} build their solution on this protocol. The key of the protocol is based on a hash function scheme, where the value of a coin \emph{C} \{0,1\} is hashed with \emph{H} to a value \emph{\~g\textsubscript{0}} in group \emph{G} of large prime order \emph{q}. This value is then raised to a secret exponent \emph{x\textsubscript{0}} to obtain \emph{\~g\textsubscript{0}}. Hash \emph{H'} exists for testing the validity of the secret share validity, and finally a hash \emph{H''} exists for restoring the coin value by calculating \emph{H''(\emph{\~g\textsubscript{0}})}.

\begin{figure}[htbp]
\begin{center}
\includegraphics[scale=.35]{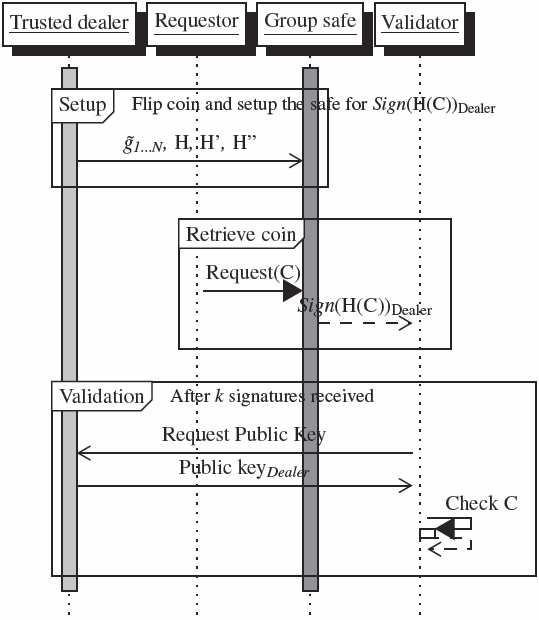}
\caption{Message exchange in the threshold coin-toss protocol, where \emph{\~g\textsubscript{1...N} = individual secret shares, H, H', H'' = Hash functions, and C = Coin}}
\label{fig:thresholdcointoss}
\end{center}
\end{figure}

The protocol message exchange is shown in figure \ref{fig:thresholdcointoss} in generalised format. An actual coin toss can be achieved if both parties are required to access the 'safe' to retrieve the result. We use this protocol in section \ref{distnoticeboard} to implement a distributed anonymous notice board. Other examples of use includes online games, where the parties must commit to a selection independently prior any party revealing their choice to others. This is useful when playing cards. Equipped with a group threshold signature, secret sharing and coin-tossing schemes, we can now proceed to the secure distributed broadcasting protocol.

%% file: distsecagreement.tex
\subsection{Secure distributed agreement}
\label{distsecagreement}
\label{distnoticeboard}
Broadcast protocols are a fundamental building block for providing replication in distributed systems. Secure asynchronous protocols are needed, for example, in distributing \emph{conference keys} among the group members, and publishing information about available services and public keys. A system where the broadcast information is stored prior to consumption, provides a distributed notice board service.

\begin{figure}[htbp]
\begin{center}
\includegraphics[scale=.35]{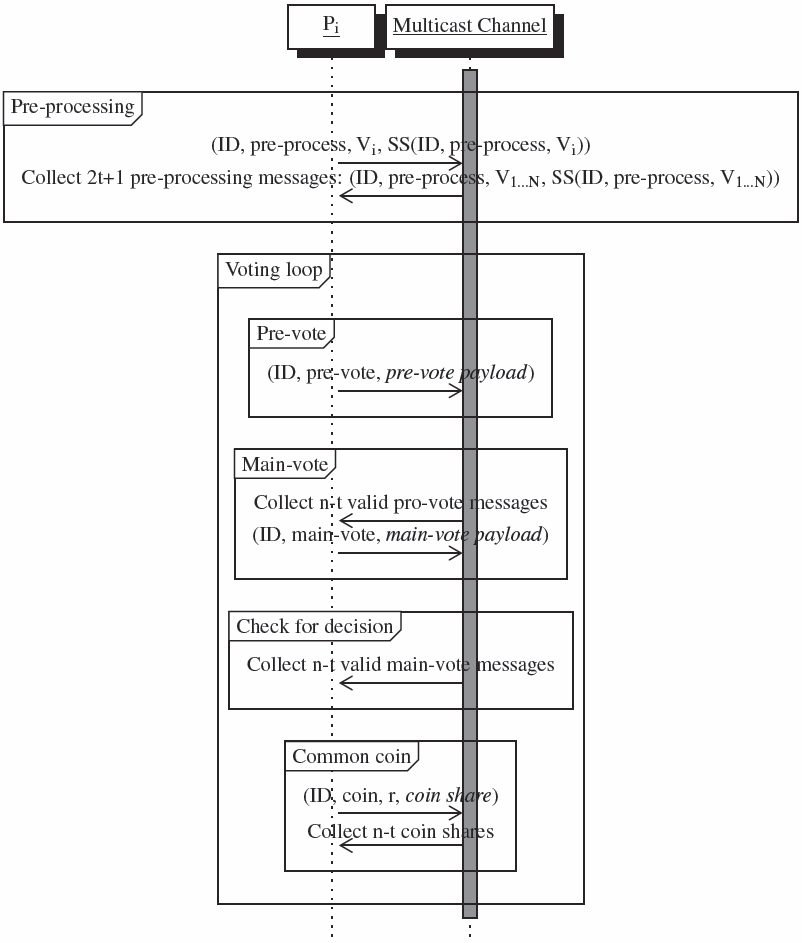}
\caption{Asynchronous Byzantine Agreement for party P\textsubscript{i}, where V\textsubscript{i} = Initial Value, SS = Signature Share}
\label{fig:byzantine}
\end{center}
\end{figure}

The concept is analogous to that of a distributed \emph{Byzantine Agreement} \cite{Lamport1982}, as the participating members must be able to agree on the correct value of particular information, store this and have a way of retrieving it afterwards without being subject to maliciously misbehaving participants trying to distribute any misinformation. We approach a solution via the work presented in \cite{Cachin2005}. Figure \ref{fig:byzantine} shows the message exchange from the perspective of one participant \emph{P\textsubscript{i}}. In the pre-processing round all participants distribute their signature share to other participants. In the following steps the parties establish what value is being voted for, and then in subsequent voting rounds look for a winning result. All steps need to meet the pre-defined thresholds in order to avoid malicious parties to collaborate. A system where the broadcast information is stored prior to consumption provides a distributed notice board service. This can be implemented with the Asynchronous Byzantine Agreement protocol by using the protocol to vote on the correctness of the information offered to a requestor by one of the data holders.

%% file: confkey.tex
\subsection{Conference key agreement}
\label{confkey}

In a conference key agreement protocol the participants together establish a common conference key to enable secure exchange of messages between the participants. Many conference key protocols have been proposed in the past with various degrees of proof for security \cite{RSALaboratories1993,Augot2005, Jung2006a, Lee2006}. The most common approach in these protocols is again based on the idea of group secret sharing (see section \ref{thresholddata}), where a trusted dealer initialises the participants with individual key shares and resolving functions. Equipped with these, the participants can each generate a portion of the conference key and share their shares between each other. Concatenation of these shares becomes the conference key.

This is, of course, subject to any malicious activity from the participants, as they could send an invalid key share to one or more participants, and in this way exclude them from the conference. Huang et al. \cite{Huanga2008} have proposed a protocol that suggests a way to filter malicious participants at the beginning of the protocol to ensure that all participants obtain the same conference key.

Conference key agreement protocols are an active research area. Recently Harn and Lin \cite{Harn2010} have proposed a protocol where a trusted dealer uses a secure broadcast, where the dealer or key generation center (KGC) broadcasts group key information directly to all participants. Another recent proposal was made by Zhao et al. \cite{Zhao2010}, where they suggest a two-stage key creation process, where the initial conference key functions as a \emph{master key} that is not used for data communications, but is used for creating \emph{'session'} conference keys to be used for the actual data exchange. Their proposal provides \emph{forward security}, which ensures that all previously issued session keys retain their secrecy even if any participant's master key would be compromised.

The conference key protocols provide the basic data confidentiality for data exchange between group participants analogously to how SSL session keys are created for one-to-one client-server connections based on the previously distributed identity certificates.

%% file: anonymousbroadcast.tex
\subsection{Anonymous broadcast}
\begin{figure}[htbp]
    \centering

\includegraphics[scale=.35]{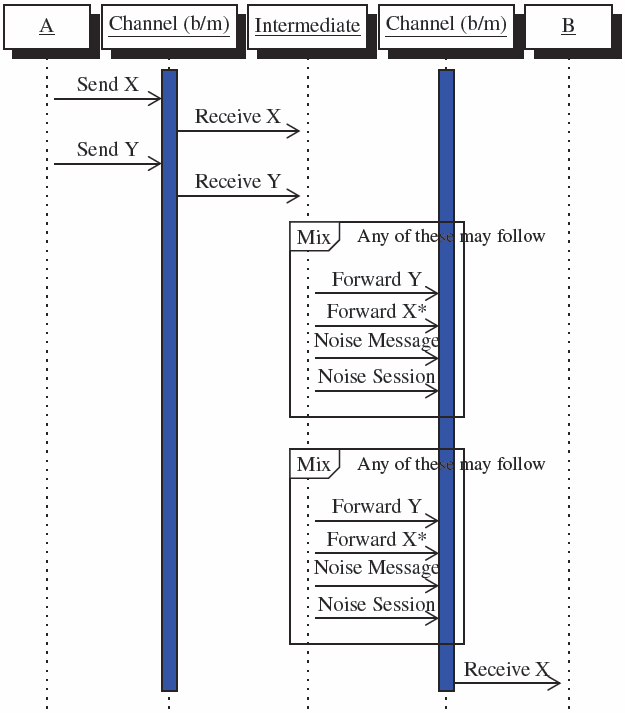}
\caption{P\textsuperscript{5}: message X forward sequence to the destination channel~(* If X\textsubscript{Sender}  $\neq$ X\textsubscript{Receiver} and m\textsubscript{B} $\leq$ m\textsubscript{A})}
\label{fig:P5}
\end{figure}

The protocol is based on P\textsuperscript{5} (Peer-to-Peer Personal Privacy) protocol \cite{Sherwood2005a}, which can be used to provide sender-,receiver-,and sender-receiver anonymity. The protocol is based on the creation of hierarchical of broadcast channels, where different levels of hierarchy provide different levels of anonymity. The channels are based on an overlay network structure, where the nodes are named based on a hash function of their real identity, and assigned to different segments of the hierarchy based on their pseudonym. Figure \ref{fig:P5} shows how the broadcast channel relays messages. The key here is that the relaying node does not know if the sender is the message originator or just another relay-step. To prevent the traceback of messages by observing the traffic in the entire network, the participants are required to send noise messages and to mix the order of received messages before relaying them.

%% file: otherprotocols.tex

\subsection{Distributed mutual exclusion protocol}
\label{otherprotocols}
The issue of allowing only a certain number of participants from accessing their critical sections simultaneously is a distributed mutual exclusion (mutex) problem. A number of distributed mutex protocols have been proposed in the literature. Recently Chaudhuri and Edward \cite{Chaudhuri2008} have proposed a protocol based on tokens and a hierarchical network model, where the participants can enter their critical section if they have first gained the possession of the token. Their approach relies on the construction of local and global mutex request queues.

In our organisational email environment, these queues could be managed by the allocated leaders of the group, assigned to a dedicated queue manager, or distributed using the group threshold signature and safe protocols. We leave the specifics of this discussion to a later work. 

%% file: distributedpatterns.tex
\section{Distribution and de-centralisation of workflow patterns}
\label{distributedpatterns}

The workflow patterns as reviewed in \cite{Aalst2003} and \cite{Russell2007} consider workflows in a centrally controlled workflow management environment. They outline 126 different patterns, and while it would be interesting to discuss all of these from the decentralisation perspective, we have selected a set of key patterns in order to discuss fundamental aspects of distributed workflow management, the application of the security protocols discussed in Section \ref{securityprotocols}, and a set of more complex patterns that highlight some of the more challenging aspects of process distribution. We discuss the control-flow, data, and resource management perspectives of the patterns, and focus on the aspects that relate to management of task concurrency and co-ordination between participants -- aspects that are normally managed by a central workflow management system, but require participant co-operation in a de-centralised setting.
In the following sections we use the pattern numbering introduced in \cite{Aalst2003} and \cite{Russell2007}.

\input{controlflowpatterns}

\input{datapatterns}

\input{resourcepatterns}

%% file: controlflowpatterns.tex
\subsection{Control-flow patterns}
\label{controlflowpatterns}

The control-flow patterns describe tasks and how their execution is controlled via connecting constructors like sequence, choice, parallel execution and join. From the perspective of de-centralisation, interesting control-flow issues relate to aspects of concurrency management, and co-ordination between the participants. We first discuss the \emph{Sequence} pattern, which provides the foundation for linking tasks into processes. The \emph{Multiple instances without a priori runtime knowledge} show how different participants can co-operate in ensuring workflow structure, and the \emph{Interleaved Parallel Routing} provides an example of a task sequence ordering between parties that are executing concurrent threads of a flow.

\subsubsection{Pattern WCP-1 (Sequence)}

The \emph{Sequence} pattern provides the most fundamental building block for processes by enabling the sequential execution of tasks one after the other. Traditionally a CWMS tracks the progress of processes and manages the allocation of the \emph{task instances} (work items) to resources (see Section \ref{resourcepatterns}), provides status information, and handles error situations. In a decentralised environment the participants need to co-operate in achieving the same functionality, for example, by using a group broadcast protocol for sending updates on process status.

All work item life cycle stages \emph{(offered, allocated, withdrawn, started, completed, failed)} and transitions between these stages need consideration in the decentralised model. Who will maintain oversight of the work item orchestration? In our proposed SOE the instance that initiated the process for the service can perform this function, and as services are built on other services, the orchestration is a hierarchical responsibility. If a truly distributed orchestration service is required, the participating members can share the work item information via secure broadcast protocols and use existing scheduling algorithms to share the management load.


\subsubsection{Pattern WCP-15 Multiple instances without a priori runtime knowledge}

\emph{"Within a given process instance, multiple instances of a task can be created. The required number of instances may depend on a number of runtime factors, including state data, resource availability and inter-process communications and is not known until the final instance has completed. Once initiated, these instances are independent of each other and run concurrently. At any time, whilst instances are running, it is possible for additional instances to be initiated. It is necessary to synchronize the instances at completion before any subsequent tasks can be triggered."} \cite{Russell2007}

\begin{figure}[htbp]
\begin{center}
\includegraphics[scale=.35]{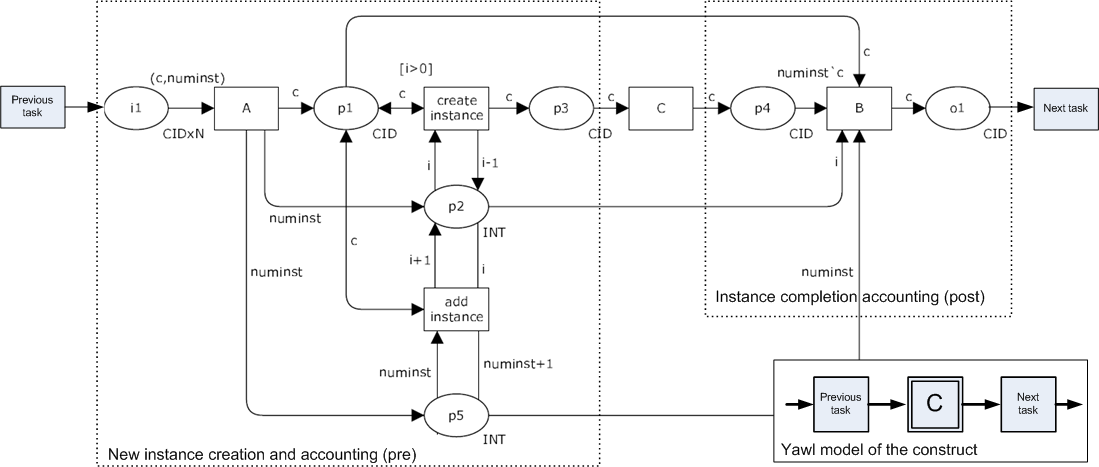}
\caption{Pattern \emph{WCP-15 Multiple instances without a priori runtime knowledge} together with its YAWL-model counterpart, adapted from a CPN-model in \cite{Russell2007}}
\label{fig:wcp15}
\end{center}
\end{figure}

Task C in figure \ref{fig:wcp15} is the multiple instance task. The two dashed areas (pre and post) show the areas of the CPN petrinet model that implement the concurrent instance creation and accounting, and the termination and merging chores respectively. While in a centrally managed workflow these chores would be performed centrally; in a decentralised setup the activity must be distributed. This chore can be split in a number of ways. For example, the pre-processing activity could be performed by a node that was allocated it by a workload balancing algorithm, and the post-processing activity could be performed by the node that is allocated with the next task. Whether this is an acceptable arrangement would depend on whether the next task can be trusted with the task of merging the parallel threads, as they would be in a position to omit individual results from selected threads. Should this be an acceptable arrangement, the pre-processor and the post-processor are required to establish a secure communication channel, for example, by using a conference key agreement protocol, so that the pre-processor can safely pass the information from \emph{p1}, \emph{p2} and \emph{p5} to the post-processor.

\subsubsection{Pattern WCP-17 Interleaved Parallel Routing}
\begin{figure}[htbp]
\begin{center}
\includegraphics[scale=.35]{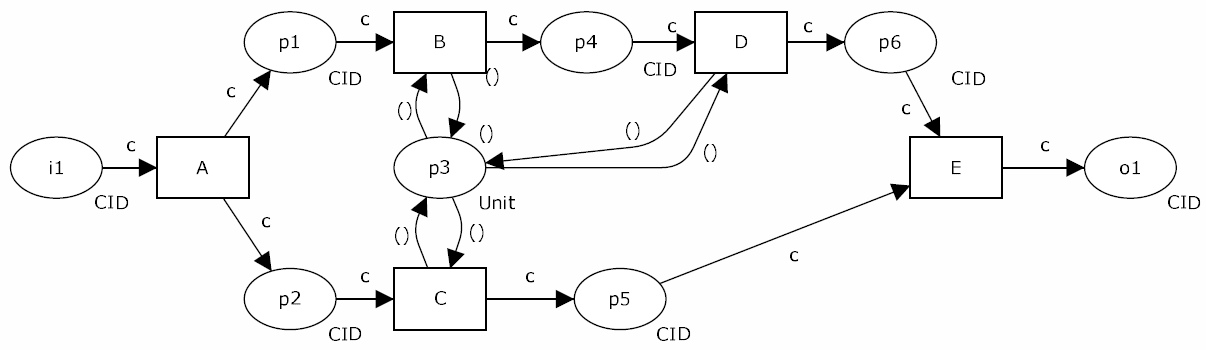}
\caption{Pattern WCP-17 Interleaved Parallel Routing as depicted in \cite{Russell2007}}
\label{fig:wcp17}
\end{center}
\end{figure}

\emph{"The Interleaved Parallel Routing pattern offers the possibility of relaxing the strict ordering that a process usually imposes over a set of tasks. Note that Interleaved Parallel Routing is related to mutual exclusion, i.e. a semaphore makes sure that tasks are not executed at the same time without enforcing a particular order."} \cite{Russell2007} In figure \ref{fig:wcp17} the place p3 implements a mutex variable in the CPN-model of the pattern. In a de-centralised environment an equivalent implementation can be achieved using the distributed mutual exclusion protocol outlined in section \ref{otherprotocols}.

%% file: datapatterns.tex
\subsection{Data patterns}
\label{datapatterns}

Data are the second key ingredient in workflow processing; the control flows create, manipulate, transfer and store data as well as change their behaviour, for example, by making routing decisions based on the data values. Each one of these aspects provides interesting challenges in a decentralised workflow environment. We follow the visibility, interaction and transfer patterns defined in \cite{Russell2007}.

\subsubsection{Data visibility patterns}

\begin{figure}[htbp]
\begin{center}
\includegraphics[scale=.18]{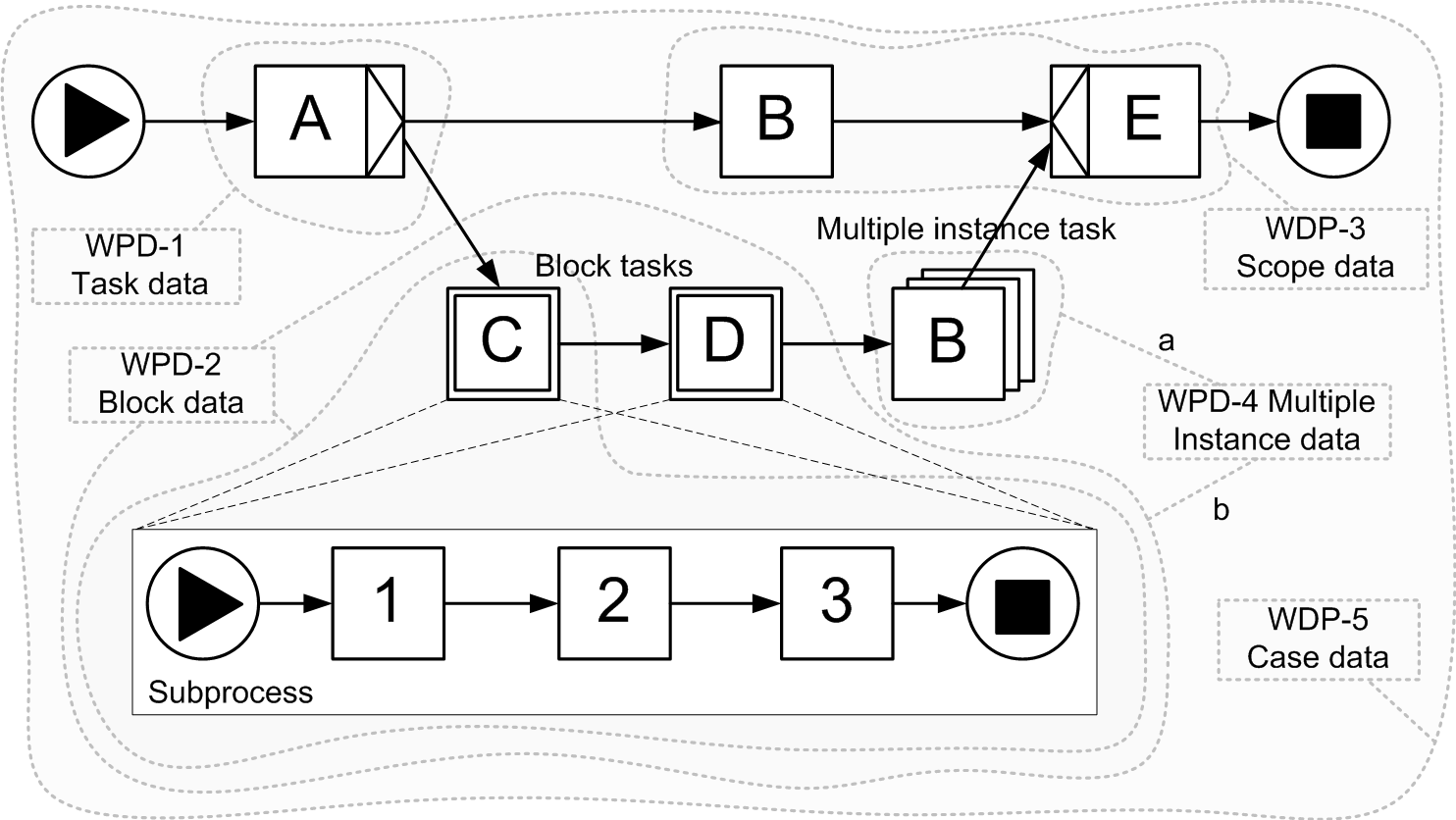}
\caption{Summary of process data visibility patterns}
\label{fig:datascope}
\end{center}
\end{figure}

The data elements can be defined and utilised at the different levels of the workflow process definition. The data can be confined to single tasks (WPD-1), accessed by all components of a sub-process (WPD-2) or by a defined set of tasks (WPD-3), shared by parallel multiple task instances (WPD-4 a) or block tasks (C \& D) that share the same sub-process implementation (WPD-4 b), or accessed by all tasks within a workflow instance (WPD-5) (Figure \ref{fig:datascope}). The data visibility scopes can be enforced by dynamically creating conference keys for the participants involved with the tasks in scope. Similarly to the data visibility scopes, group signature and anonymous broadcast channels can be dynamically created for the participants where these services are needed.

\subsection{Data interaction patterns and transfer patterns}

Each of the data visibility patterns need to be supported by a mechanism to communicate the data elements between process components. Generally speaking, the data interaction can take place from task to task, from block task to sub-process and back, from task to multiple instance task and back, from process instances to instance, and from the external environment to process components and back. The data transfer mechanism in these different forms of interaction; can happen a number of ways, namely by moving or copying the data by value, or by reference with or without locking source location of the element.

In the case of the \emph{SOE}, the data elements are transferred by email messages. Most of the messages can be handled automatically, with only the messages that require user actions displayed. However, the transfer of data elements does not need to follow the control flow of the workflow. For example, in the case of the WPD-2 pattern, \emph{Task C} could pass all of the elements that \emph{Tasks 1, 2 and 3} require to \emph{Task 1}, which could then forward elements to \emph{Tasks 2 and 3} via \emph{Task 2}. Should some data not be required by and be kept confidential from \emph{Tasks 1} and \emph{2}, this could be encrypted with a shared key between \emph{Tasks C} and \emph{3}, but \emph{Tasks 1} or \emph{2} could still maliciously interfere with the data. Alternatively, \emph{Task C} could send these data directly to \emph{Task 3}.

\begin{figure}[htbp]
\begin{center}
\includegraphics[scale=.20]{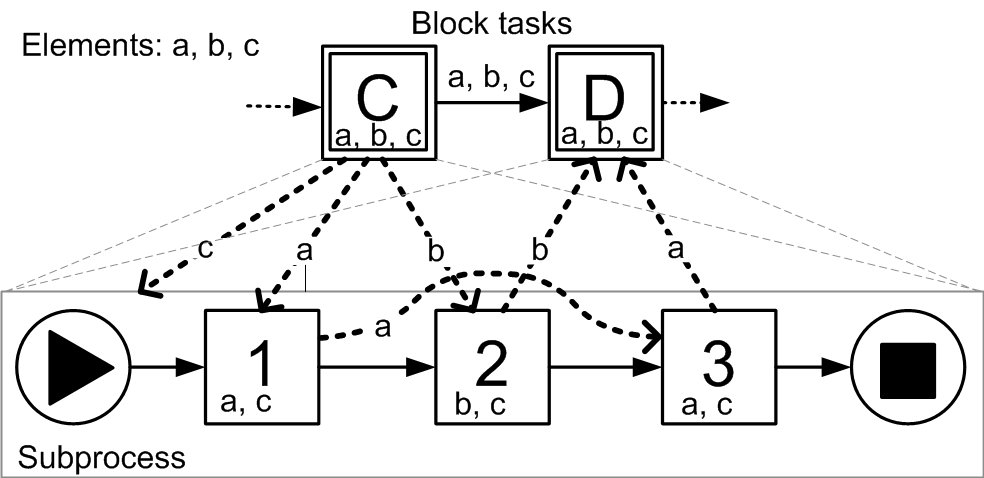}
\caption{Data-flow semantics as part of a workflow model}
\label{fig:dataflow}
\end{center}
\end{figure}

Obviously, the most suitable approach for moving the data depends on the nature of the data and the trust levels and security clustering of the participating tasks. This demonstrates how the data-flow design must be designed in parallel with the controlflow design, and supported by integrated data-flow and security modeling tools. Figure \ref{fig:dataflow} shows data flow behavior with control flow.

%% file: resourcepatterns.tex
\subsection{Resource patterns}
\label{resourcepatterns}

Resources that perform the tasks in the workflow processes are the third key ingredient to complete workflow definitions. Patterns from WRP-1 to WRP-39 \cite{Russell2007} are also all feasible usage structures in decentralised processing environments. We discuss below three patterns that are relevant to our earlier discussion on security protocols.

\subsubsection{Pattern WRP-2 Role-based distribution}
\emph{The ability to specify at design time one or more roles to which instances of this task will be distributed at runtime. Roles serve as a means of grouping resources with similar characteristics.} \cite{Russell2007} The role-based task allocation goes hand in hand with the data visibility and transfer patterns when enforced with data encryption. For example, the role holders will also need to be assigned with the corresponding conference or secure broadcast channel keys.

\begin{figure}[htbp]
\begin{center}
\includegraphics[scale=.20]{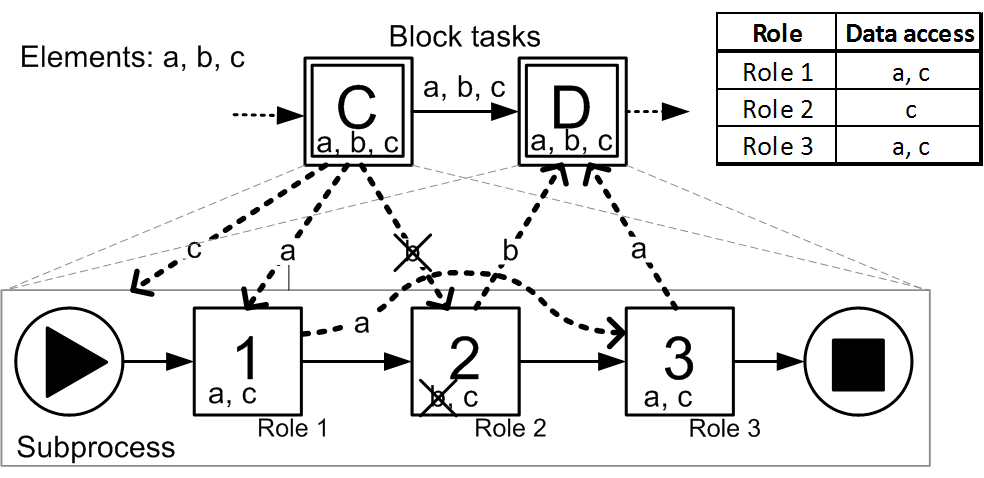}
\caption{Role data with Control and Data flow semantics as part of a workflow model}
\label{fig:dataflowrole}
\end{center}
\end{figure}

Combined with the control-flow and data-flow modelling, task role allocation modeling enables us to analyse and show inconsistencies between these different aspects of model definition (figure \ref{fig:dataflowrole}). In a decentralised workflow environment the interplay between the required data elements, allocated resources, and available new security services is complex, and requires integrated modelling tools for process design and monitoring.

\subsubsection{Pattern WRP-13 Distribution by Offer - Multiple Resource}

While the original WRP-13 pattern refers to a situation where a work item is offered to a number of suitable resources, of which one will accept the offer, the pattern could be extended to allow a group of resources to accept an offer to perform a block task. Considering the threshold group signature scheme (discussed in \ref{thresholdsignatures}), the role-to-resource mapping should not only be considered as a way of grouping resources with similar characteristics. To return to our earlier example where the ability to sign a document depended on how many of the required signature shares can be pooled together by a group of users, this could introduce a different type of \emph{team-role} pattern, where a block task can be allocated to a team of resources who together can perform the activities within the sub-process.

%% file: distributedwfsim.tex
\subsection{About our reference model implementation}
\label{distributedwfsim}

The YAWL project \cite{Aalst2005} has produced a detailed CPN-tools based \cite{Ratzer2003} simulation model for the \emph{newYAWL}-workflow modeling language. Our research builds on this model. This includes a number of challenges. Firstly, the main workflow management framework must support the distribution of the \emph{work distribution} function and the implementation of an asynchronous data distribution mechanism for shared variables required for the overall management of the system. We aim to implement minimal core functionality in the CPN-model, and the security, data distribution and mutual exclusion protocols as workflow definitions.
